\documentclass[aps,prc,amsmath,amssymb]{revtex4}
\usepackage{graphicx,amsmath,bm}

\begin{document}
\title{Luge Track Safety}
\author{Mont Hubbard}
\affiliation{Department of Mechanical and Aerospace Engineering, University of California, Davis, Ca., 95616 USA}
\date{\today}

\begin{abstract}
Simple geometric models of ice surface shape and equations of motion of objects on these surfaces can be used to explain ejection of sliders from ice tracks.  Simulations using these can be used to explain why certain design features can be viewed as proximate causes of ejection from the track and hence design flaws. This paper studies the interaction of a particle model for the luge sled (or its right runner) with the ice fillet commonly connecting inside vertical walls and the flat track bottom. A numerical example analyzes the 2010 luge accident at the Vancouver Olympics. It shows that this runner-fillet interaction, and specifically the fillet's positive curvature up the inside wall,  can cause a vertical velocity more than sufficient to clear the outside exit wall. In addition its negative curvature along the track, together with large vertical velocity, explains loss of fillet or wall contact and slider ejection. This exposes the fillet along inside walls as a track design flaw. A more transparent design and review process could provide a wider, more complete and thorough design and verification scrutiny by competent scientists and engineers without financial involvement or conflicts of interest and potentially lead to safer future designs.  
\end{abstract}
\maketitle

\section{Introduction}  %

One day before the opening of the 2010 Winter Olympic Games in Vancouver, a tragic crash in the 6th and last  training run claimed the life of Georgian luge slider Nodar Kumaritashvili~\cite{FIL2010}.  The accident cast a pall over the impending Olympics and called into question whether one of the centerpieces of the Games, the ice sliding track, was safe enough to proceed. 

A series of investigations and reviews  followed immediately. Television videos of the accident were analyzed by the International Luge Federation (FIL) whose representatives almost immediately declared that the athlete was responsible and that the accident was caused by a steering error rather than a track design flaw. Yet several changes were immediately made to the track including:
1. increasing the height of the outside wall by 2.26 m and 1 m for 18 m and 28 m, respectively,
2. increasing the height of the inside wall by 0.4 m for 46 m, and
3. "Squaring off the curve of the ice between the base of the track and the side walls of the outrun"\cite{FIL2010}.
All the Olympic luge events were eventually contested from the lower Women's start to limit top speeds.

\begin{figure}
\hskip .2 in\includegraphics[width=.25\textwidth]{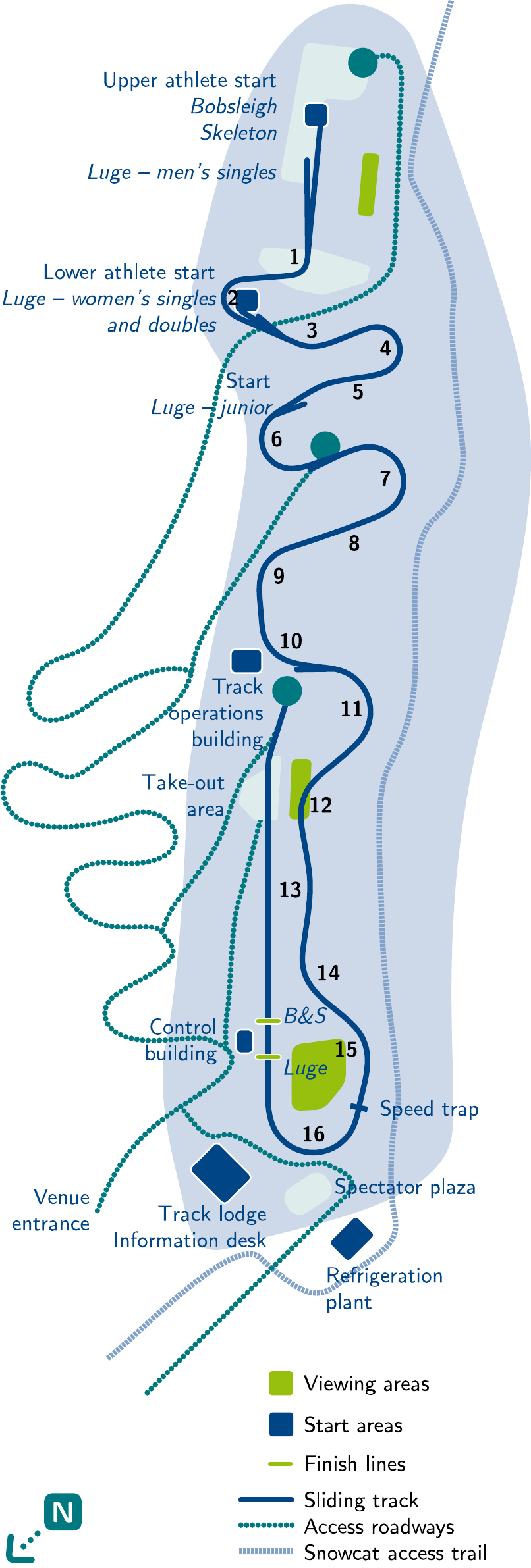}
\caption{Plan view of Whistler ice track. The accident occurred at the very bottom of the track on exit from Turn 16 at a speed of 39.81 m/s. } 
\label{WhistlerTrack} 
\end{figure}

\begin{figure}
\hskip .2 in\includegraphics[width=1\textwidth]{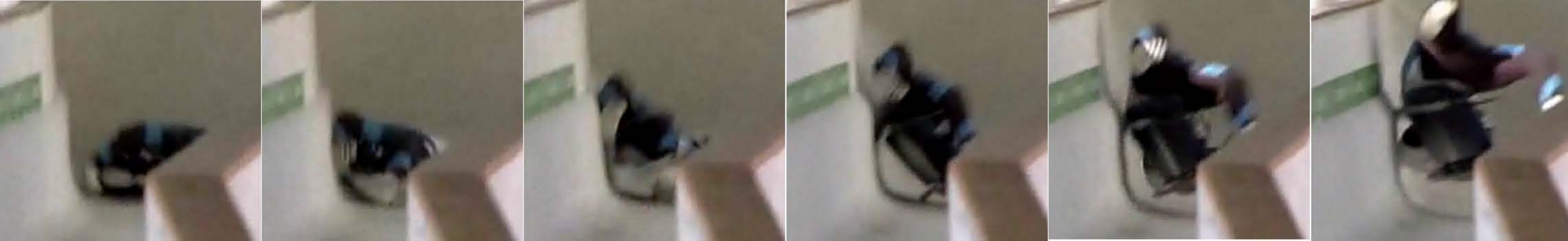}
\caption{Frames of accident (referred to as a-f from left to right) looking back into Turn 16 spaced at 0.033 s. Contact of right  (left as viewed) runner with the inner wall lasts considerably less than 0.1  s. (Photos captured by Les Schaffer.)} 
\label{StillFrames} 
\end{figure}

The accident occurred at the bottom of the track near the exit from the last turn (Turn 16 in Fig.~\ref{WhistlerTrack}) at a near maximum speed of 143.3 kph (39.8 m/s) and was captured in real time with video footage. Several consecutive frames looking at the oncoming sled are shown in Fig.~\ref{StillFrames}.  Later analysis~\cite{FIL2010} concluded that, reacting to perturbations in the previous Turn 15 and to contact of the slider's right glove with the ice during Turn 16, the sled descended the banked portion at the exit of Turn 16 too early, and crossed the flat bottom portion of the track (Fig.~\ref{StillFrames}a).  Its right runner then contacted the lower portion of the inner wall  (Fig.~\ref{StillFrames}b), rode up the inner cylindrical wall Fig.~\ref{StillFrames}c, d while the left runner continued touching the floor.

Since the accident several analyses of it  have been published. Within two months an Official Report by FIL~\cite{FIL2010} appeared. Most of this dealt with  athlete qualification procedures, track construction and homologation (certification),  and previous crash histories. Although the report at least recognized that this ``tragic result ... should not have occurred as a result of an initial driving error"~\cite{FIL2010}, the few sentences dealing with the accident itself offered only a pseudo-scientific explanation: the sled ``appears to have hit the wall at an exceptional angle that caused the sled to compress..result[ing] in the sled serving as a catapult when it decompressed launching ... the sled into the air"~\cite{FIL2010}. Furthermore it classified the dynamics of the crash as ``unknown and unpredictable"~\cite{FIL2010}.

A Coroner's report~\cite{Coroner2010} followed after seven months. It found that there was ``no evidence that any serious injury resulted from the initial impact with the inside wall of the track", and it at least correctly recognized  that ``the collision with the inner wall caused the sled's right runner to ride up onto the wall, causing ... the sled to be launched into the air."~\cite{Coroner2010}.  It also concluded, qualitatively,  that ``The bottom inside corner ... had a rounded profile that influenced the rate and angle at which the sled runner climbed the wall, ultimately affecting the trajectory of [the] ejection"~\cite{Coroner2010}. This report recommended an independent and comprehensive safety audit of the track.

The vertical velocity that resulted in track ejection was acquired entirely during this launch process that occupied at most three frames (less than 0.1 s). Yet nearly two years after the incident, no investigation has clearly described how the vertical velocity required for the rider to be ejected from the track, i.e. to clear the outside wall at the exit of the turn, was acquired.  The purpose of this paper is to provide the first actual explanation, based on physical principles,  of how the sled and slider were ejected from the track and to explain the track design features responsible. It focuses on the rounded shape of the bottom inside corner. In the remainder of the paper this is called the ``fillet". Finally the paper further examines the design and review process as it exists today and makes suggestions to increase track design safety. 

\section{Methods} 

\subsection{Sled Description}

Before proceeding it should be asked, "What do luge sleds look like?" with a view to whether a particle approximation for its (or its right blade's) motion is a reasonable one. The two runners, although not sharpened like ice skate blades, are thin enough that they strongly indent the ice surface. The sled goes where the runners are pointed, and there is little or no lateral slip of the runners on the ice.  Although the lateral cross section of the steel runner surface contacting the ice is not exactly circular or even uniform, it suffices to think of them as a surface with major (longitudinal) radius $R_s \simeq$ 13 m and minor (lateral) radius $r_s \simeq$ 0.003 m. The steel runner-ice  contact patch length has been estimated to be of the order of 12 cm. Luge rules specify that the runner ice contact points may not be separated laterally by more than 0.45 m~\cite{FIL2011}.

\subsection{Ice Track Description}

\begin{figure}
\footnotesize
\begin{minipage}[c]{0.45\textwidth}
                \centering
                \includegraphics*[width=.9\textwidth]{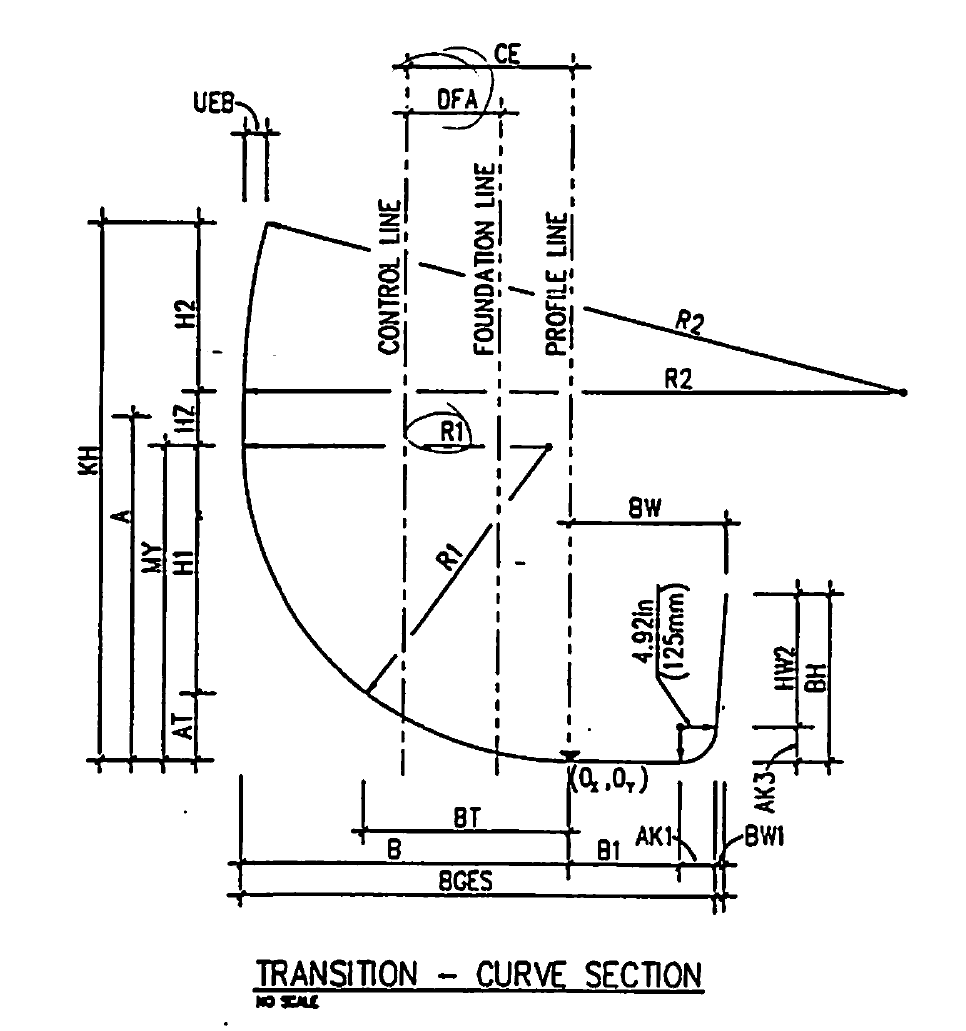}

                (a)
\end{minipage}
\begin{minipage}[c]{0.45\textwidth}
                \centering
                \includegraphics*[width=.9\textwidth]{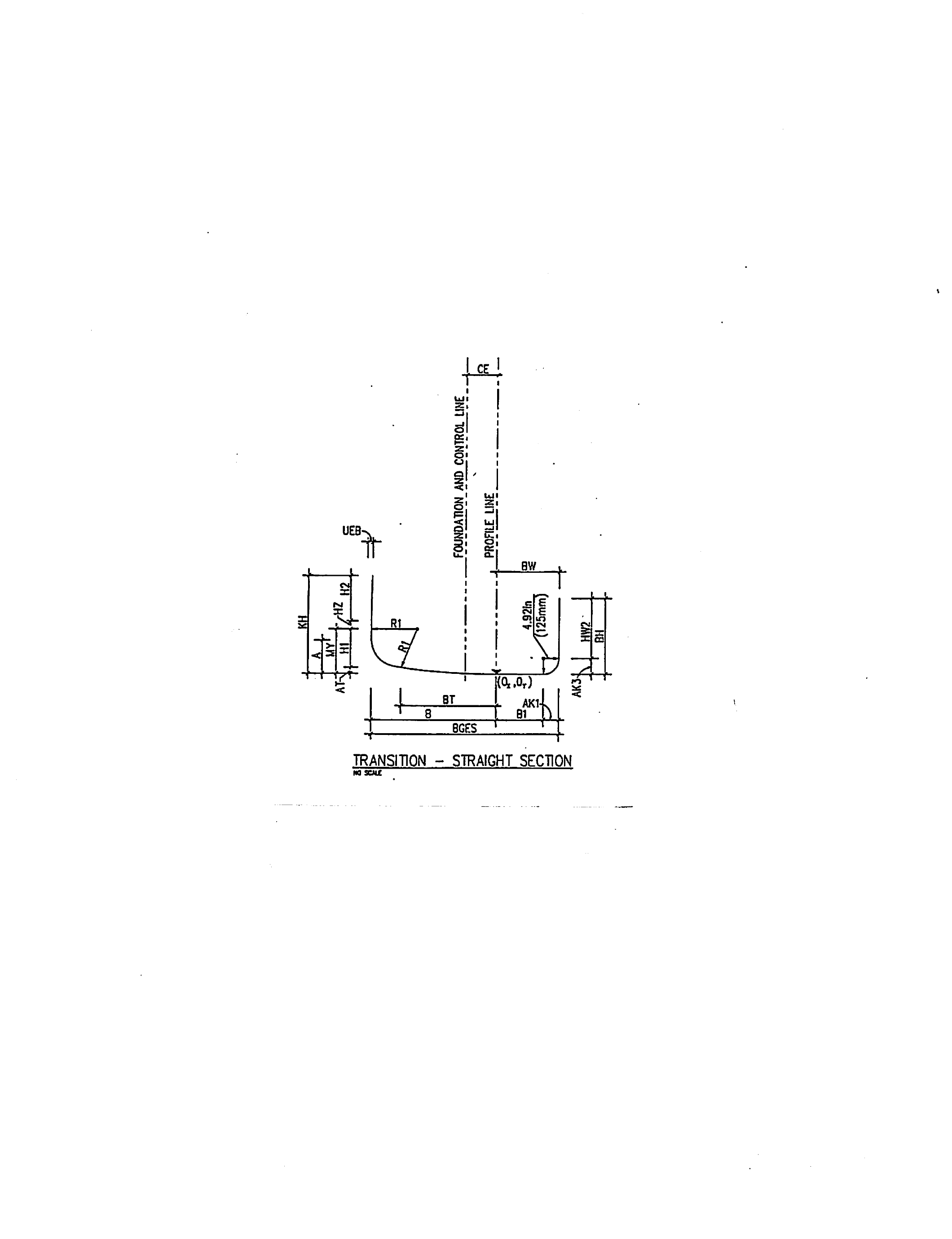}

                (b)
\end{minipage}
\caption{Ice track cross sections are strongly banked in turns and roughly rectangular in straight segments. Shown here are two example cross sections from a banked turn on the Calgary track; a) near the turn center, and b) near the turn exit where the cross section has become nearly rectangular. Both the curved and nearly straight cross sections contain a small radius ($r=125$ mm) concave fillet at the lower right corner connecting the flat bottom to the right (looking downtrack) inside vertical side wall. Seven of the last 8 Olympic tracks have been created by the same design firm, IBG of Leipzig, Germany and each has this design feature.}
\label{crosssections}
\end{figure}

Ice tracks are specific to the hill on which they are constructed. A particular track centerline profile is laid out on the chosen hill to have the correct average slope in order to give the gradual  buildup and maintenance of speed desired. Tracks consist essentially of straight segments connected by strongly banked curves (see Fig.~\ref{WhistlerTrack}). In straight segments the cross section consists of two exactly vertical walls separated by a flat bottom (floor) about 1.5 m wide (close to the shape shown in Fig.~\ref{crosssections}b). A typical banked cross section nearer the middle of the turn is shown in  Fig.~\ref{crosssections}a. The inside walls of the banked sections in curves are also vertical(Fig.~\ref{crosssections}a and b) and are adequately approximated locally as surfaces of right circular cylinders. 

Because vehicle transitions from the curved into the straight sections are not always centered and/or parallel to the center line, the vertical walls in the straight sections  are joined to the flat bottom with a small radius (0.125 m) concave ice fillet, described as a "bottom inside corner ...[with] a rounded profile" in~\cite{Coroner2010}. This is clearly visible in design drawings in Fig.~\ref{crosssections}b and in Fig.~\ref{StillFrames}, the shape made plainly visible there by shadows. Indeed this fillet is required in section 16.14 Straights of the International Bobsled Federation (FIBT) rules~\cite{FIBT2011} which states that "The transition between the sidewall and the base of the track must be provided with a channel [what is referred to here as a fillet]. In the iced state its radius must be 10 cm." Presumably this fillet allows bobsled (and luge) runners to gently interact with the fillet, providing a centering action in the otherwise flat straight sections. 

\normalsize

Although this rationale justifies the existence of the fillet in straights, the author is unaware of any such justification for use at the base of inner walls in banked sections. Section 16.11 Bends of the FIBT rules~\cite{FIBT2011} is notable for the absence of a similar specification. Nevertheless the fillet is also included in the design drawings of banked cross sections (Fig.~\ref{crosssections}a) connecting the inner wall to the floor.

The surface of the fillet itself can be approximated by the lower-inner one fourth of a large aspect ratio $A=R/r$ torus with major radius $R$ roughly equal to the turn radius, and minor radius $r$ equal to the fillet radius, with the surface tilted in general from vertical through the average track slope at the particular turn location to allow connection with the gradually sloping floor.

Even though Fig.~\ref{StillFrames} shows clearly that there were two points of contact during the critical period of the accident, it is not necessary to invoke a two-point contact model  to understand the essential mechanism of the sled interaction with the ice fillet  surface. The simplest model with the ability to produce descriptive results is a particle model in which a point mass slides along the surface~\cite{Hubbard1989}. Models like these have previously been used to describe the dynamic motion of bobsleds on similar ice tracks~\cite{Hubbard1989},~\cite{Kelly2000},~\cite{Zhang1995}.  

The geometry of the torus is generally described parametrically in a reference frame with origin at its center. Consider a similar $x_1y_1z_1$ reference frame with origin $O_1$ on a vertical axis at the center of the turn, $z_1$ vertical, the $y_1$ axis toward the banked track surface, but with $O_1$ translated downward from the torus center along $z_1$ by $r$ so that $y_1$ passes through the point $O$ of first contact of the particle sled with the bottom of the fillet. The equation of a local approximation of the cylindrical inner wall in this frame is 
\begin{equation}
x_1^2+y_1^2= (R-r)^2.\label{wall}
\end{equation}
where $R$ and $r$ are the major and minor torus radii, respectively, and the location of point $O$ in this reference frame is $(0,R,0)$. 

Now define the primary reference frame $xyz$ with origin at $O$ as a rotation of frame $x_1y_1z_1$ about $y_1$ through angle $\beta$, the slope of the floor relative to horizontal,  followed by a translation of $R$ along $ y_1$  so that $x$ is tangent, and $z$ is perpendicular, to the sloping helicoidal floor at point $O$.  The position vector $\mathbf{r}$ from $O$ to a general point $P$ on the toroidal approximation of the fillet in frame $xyz$ is written as
\begin{equation}
\mathbf{r}= (x ,  y , z)
\label{positionvector}
\end{equation}
where it is assumed the difference between the position vector $\mathbf r$ and scalar minor torus radius $r$ is clear from context.

Parametrically the coordinates of a point $P$ on the surface can be written as 
\begin{align}
x(u,v) = (R-r\sin v)\sin u   \notag \\
y (u,v)= R(\cos u -1)-r\sin v\cos u   \\
z(u,v) = r(1-\cos v)   \notag
\label{fillet}
\end{align}
where the parameters $u$ and $v$ are angles along- and cross-track measured, respectively, about $z$ at the turn center and about the axis of the toroidal tube. Parameter $v$ is the angle between the vertical and the line in a vertical plane perpendicular to the circular tube centerline from a point $C$ on the tube centerline to the point $P$ ($v$ is zero when $P$ is directly below $C$). Intuitively the angles $u$ and $v$ are the  ``longitude" and ``latitude"  of $P$, respectively, and the entire fillet ice surface lies in the range $0 < v < \pi/2$. Equations 3 make it clear that the position vector to an arbitrary point on the torus is a function $\mathbf{r}=\mathbf{r}(u,v)$ only of the two angles that parameterize the fillet surface~\cite{Faux1979}. Appendix A discusses the geometry of the toroidal fillet surface and presents the two equations of motion (coupled second order ordinary differential equations for the two parameters) of  a particle sliding on the surface. These describe how $u$ and $v$ must change in accordance with Newton's laws.

\section{Results} 

The equations of motion were used to study the interaction of the sled  with the fillet surface. A MATLAB simulation program was created to solve the initial value problem to calculate $u(t)$ and $v(t)$ from the initial conditions at point $O$, the point of first contact with the fillet where $u(0)=v(0)=0$, and with initial speed $s_o$ = 39.81 m/s at a variable \textit{entry angle} $\gamma_o$ relative to the $x$ axis. Thus the velocity initial conditions are $\dot x(0)= s_o\cos \gamma_o$ and $\dot y(0)= -s_o\sin \gamma_o$ which correspond to initial conditions for the speed parameters $\dot u(0)= s_o\cos \gamma_o/R$ and $\dot v(0)= s_o\sin \gamma_o/r$.

Physical and track geometric parameters used are shown in Table~\ref{Table1}. The combined slider and luge mass $m$ is realistic, but the dynamic equations of motion Eq.~\ref{EOM} are  mass independent (see Appendix A). Thus a precise value of $m$ is not important. Present day track centerline layout designs use clothoid curves that linearly vary the track centerline curvature to smooth transitions at inlets to and outlets from curves. So even though the minimum radius of curvature in Turn 16 is about 33 m, at the location of the inner wall collision $R$ was closer to 38 m as measured from track overhead views. Because the accident occurred at the very bottom of the track near where the average 0.105 downgrade reverts to an uphill one to assist in sled deceleration, it is assumed that the centerline slope $\beta = 0$. Although the speed in the Whistler accident was measured to 1 mm/s in real time, the top speed during qualifying was at least 153.0 Km/h (42.78 m/s)~\cite{Coroner2010} and this speed should be used in further safety predictions for this track. The exact value of $h_w$ is also not too important since, as will be shown below, the height achieved in flight $h_z$ is a linearly increasing function of the angle $\gamma_o$ at which entry to the fillet occurs. 
\begin{center}
\begin{table*}
\caption{Physical and Geometric Parameters}
\begin{tabular}{lccr}
\hline\noalign{\smallskip}
Parameter &Symbol& Units& Value\\
\noalign{\smallskip}\hline\noalign{\smallskip}
Acceleration of gravity & $g$   &   m/s$^2$   &   9.81  \\
Mass of slider, ballast and sled    &  $m$  & kg   & 110\footnote{estimated~\cite{FIL2011}} \\
Turn radius & $R$   &   m   &   38\footnote{measured from Googlemaps} \\
Fillet radius    &  $r$  & m   & 0.125\footnote{see Fig.~\ref{crosssections}a} \\
Centerline slope &  $\beta$   &   rad     &   0\footnote{estimated}   \\
Initial speed&  $s_o$   &   m/s     &  39.8   \\
Exit outside wall height & $h_w$& m & 1.0\footnote{estimated from photographs}   \\
Diagonal cross track distance& $d$&m& 23.4 \footnote{measured from Googlemaps} \\
\noalign{\smallskip}\hline
\label{Table1}  
\end{tabular}
\end{table*}
\end{center}
Considerable effort was made by this author (see Discussion section below) to obtain more precise values for certain parameters ($R, r, h_w$ and $\beta$) in Table~\ref{Table1} but this attempt was unsuccessful. The author believes that, while more accurate values of these parameters are somewhat desirable, additional accuracy would add very little to the conclusions of this study, and  that certainly the unwillingness of the authorities to provide these should not be allowed to prevent or further delay the analysis.

\begin{figure}
\footnotesize
\begin{minipage}[c]{0.48\textwidth}
                \centering
                \includegraphics*[width=1\textwidth]{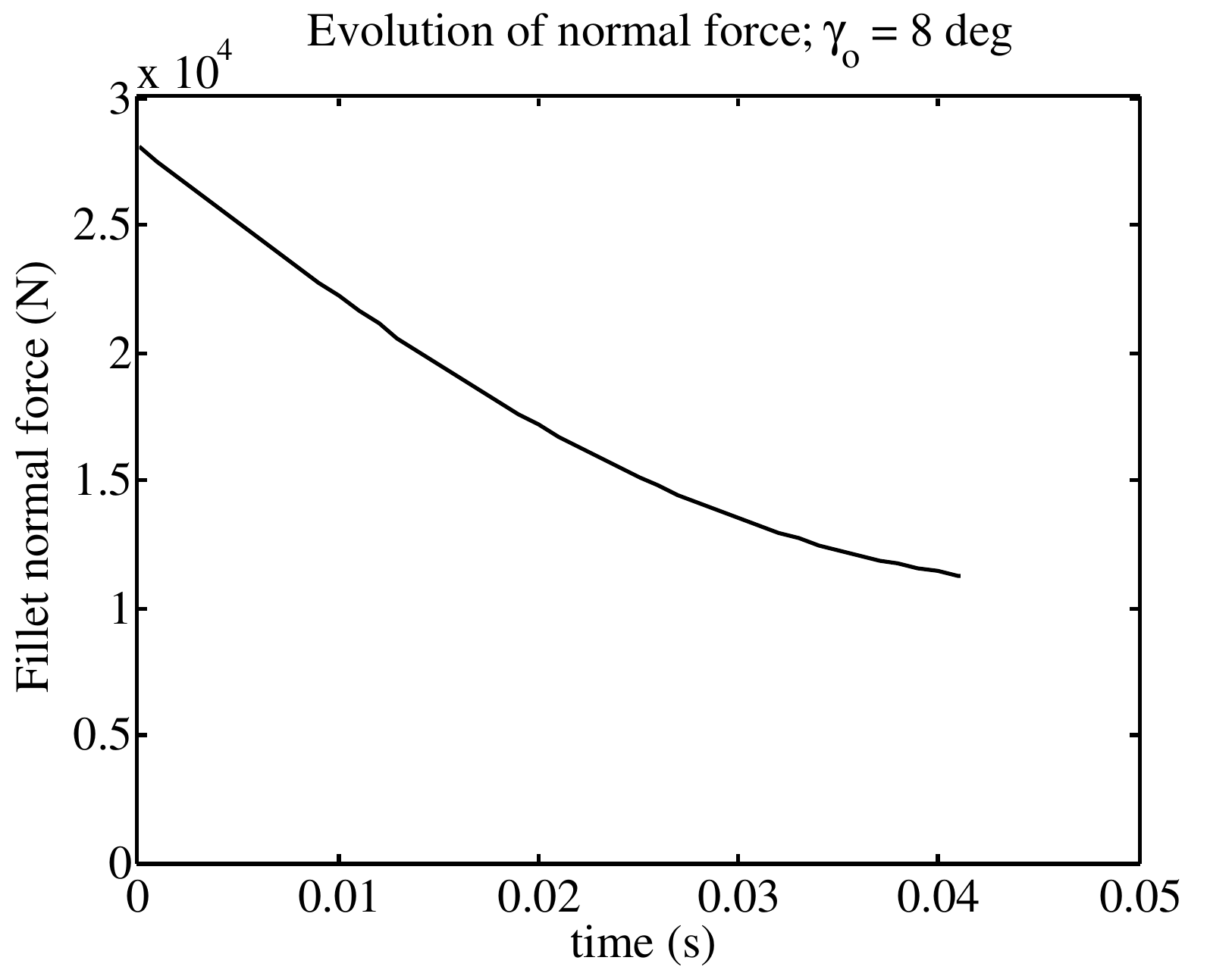}
(a)
\end{minipage}
\begin{minipage}[c]{0.48\textwidth}
                \centering
                \includegraphics*[width=1\textwidth]{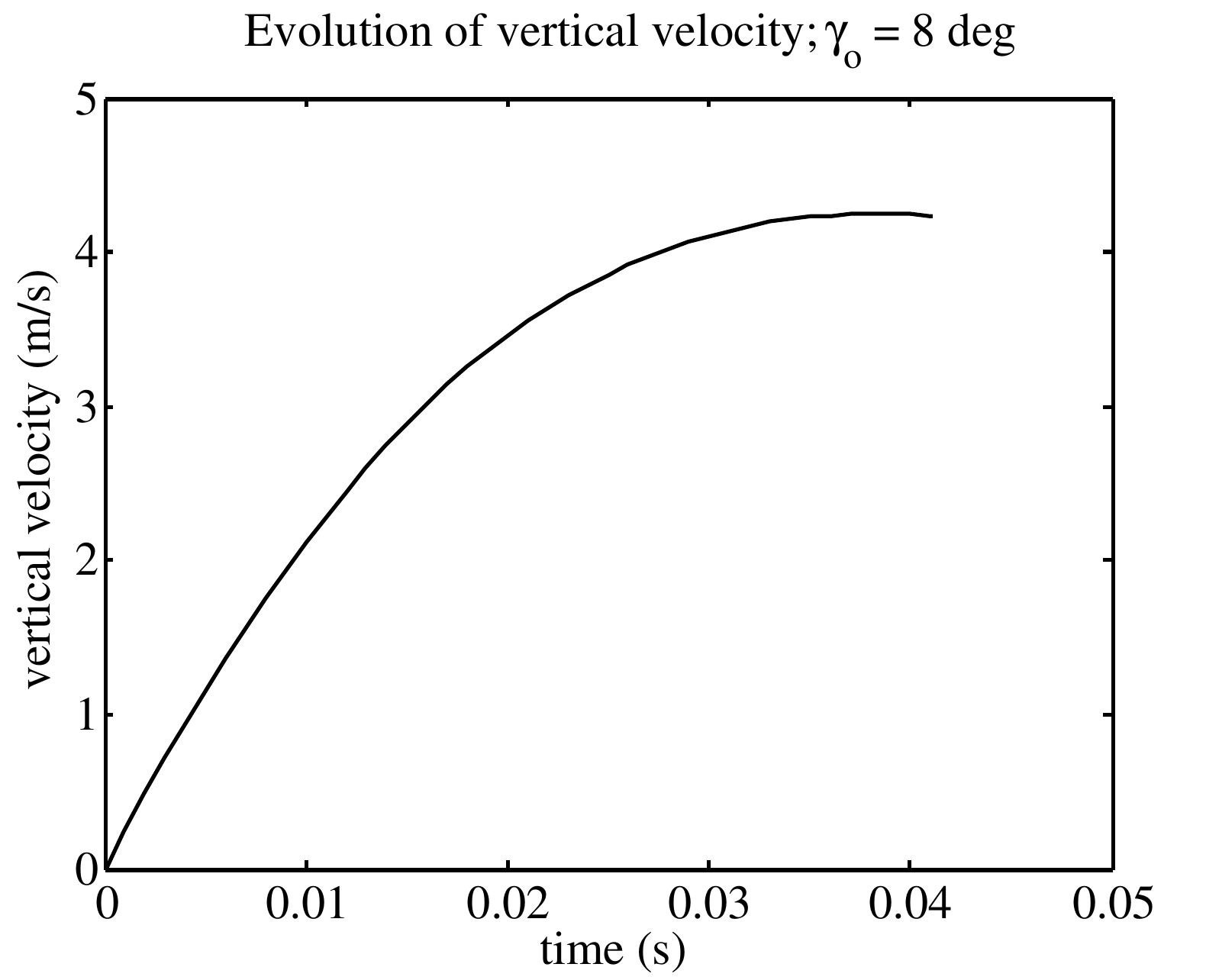}
                (b)
\end{minipage}
\caption{Simulation results for entry angle $\gamma_o = 8^\circ$.  Contact duration is only 0.041 s but contact persists until the top of the fillet is reached. a) Normal force decreases from 28 KN to 11 KN,  and the average force of $14 KN$ during the last half of contact acts nearly horizontally.  b) A substantial portion ($>80\%$) of the vertical velocity at launch is achieved in less than half the contact time because the normal force acts mostly in the vertical direction during this period.}
\label{simresults}
\end{figure}

\begin{figure}
\vskip 0.01 in
\includegraphics[width=0.4\textwidth]{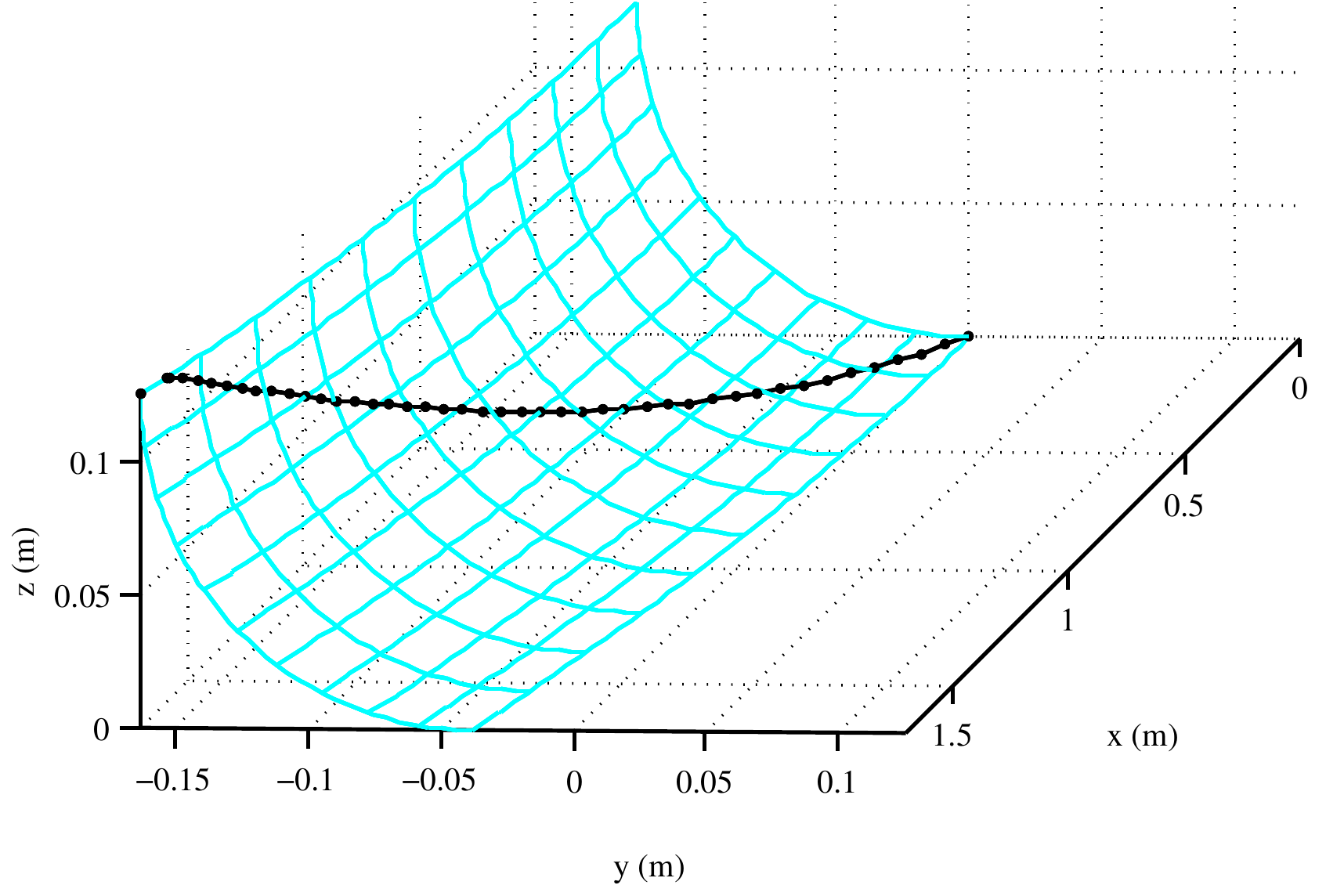}
\vskip 0.01 in
\caption{Simulation results ($x$ axis not to scale) for entry angle $\gamma_o = 8^\circ$. Luge 3D path rises to  top of fillet as inner wall turns away from initial direction. Loss of contact occurs when the positive normal curvature vanishes at top of fillet where it meets the cylindrical inner wall.} 
\label{3Dcurve} 
\end{figure}

Results from a simulation with 
$\gamma_o$= 8$^\circ$ (0.14 rad) (chosen as an illustrative example) are shown in Figs.~\ref{simresults} and~\ref{3Dcurve}. Total contact duration is very short, $t_c=0.041$ s. Although not shown, fillet latitude rises almost linearly in time, reaching a value of $\pi/2$ at $t=t_c$ . Figure~\ref{simresults}a  portrays the evolution of the normal force as a function of time. Because of the discontinuity in normal curvature (see Appendix A) between the flat floor ($\kappa_n = 0$) and the fillet, the normal force rises discontinuously at $t = 0$ by roughly a factor of 25 from the weight $mg$=1079 N to about 28 KN.  During the entire contact period the normal curvature in the direction of the path is positive in the sense that, along the path, the fillet surface is ``rising" to meet the velocity vector. When the luge reaches the top of the fillet ($z$ = 0.125 m) at its junction with the cylindrical inner wall, however, the wall surface normal curvature becomes negative (the two principal curvatures for a cylinder (see Appendix A) are $\kappa_1=0 $ m$^{-1}$ and $\kappa_2=-1/(R-r) $ m$^{-1}$). Contact is lost (the sled and slider are launched) because there is no component of the weight perpendicular to the vertical surface to supply the needed normal acceleration and maintain contact. The normal force decreases by more than half during the contact period but is still very large (11.5 KN) just before contact ends. 

A large fraction of the vertical speed at launch $s_{lz}$ of 4.24 m/s is generated during the first 0.02 s of contact (Fig.~\ref{simresults}b). This vertical velocity  results in a peak (zenith) height $h_z$ of the center of mass in subsequent free flight of 
\begin{equation}
h_z=0.125 +s_{lz}^2/2g .
\label{zenithheight}
\end{equation}
For this example $h_z= 1.04$ m, as high as or higher than the estimate of the retaining wall height $h_w$ at the curve exit. Larger entry angles $\gamma_o$ result in even larger vertical velocities when wall contact ends. 
\normalsize

The corresponding three dimensional path on the fillet surface is shown in Fig.~\ref{3Dcurve}. Fillet contact occurs only over a relatively short length ($\simeq$1.6 m) of track. Because, as the sled rises on the fillet, the surface itself is turning away from the down track direction, not all the initial lateral ($y$) velocity $s_o \sin \gamma_o$ can be turned (literally) into vertical velocity $s_{lz}$, as is shown in more detail below.  

\begin{figure}
\vskip 0.01 in
\includegraphics[width=0.5\textwidth]{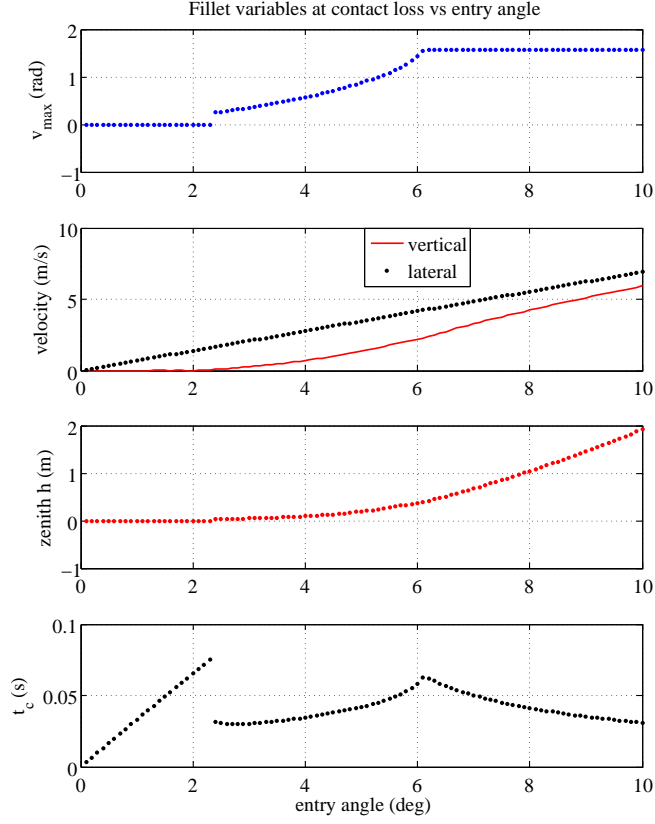}
\vskip 0.01 in
\caption{Fillet interaction variables vs. entry angle $\gamma_o$ at constant  initial speed ($s_o=39.81$ m/s) : a) fillet latitude $v_{max}$ at loss of contact, b) lateral entry speed $s_{oy}$ and vertical velocity at loss of contact $s_{lz}$ , c)  zenith height of resulting ballistic path, d) fillet contact time. Three regions of distinct behavior exist depending on entry angle. For $\gamma_o> 6.2^\circ $ contact persists continuously to top of fillet ($v_{max}=\pi/2$). As entry angle approaches $\gamma_o = 10^\circ $ almost all lateral entry velocity is converted to vertical ejection velocity and zenith height $h_z$ approaches 2 m. Distance to subsequent flight  zenith is roughly initial speed times vertical velocity divided by $g$. } 
\label{summary} 
\end{figure}

Figure~\ref{summary} shows  a summary of results of 100 such simulations for entry angles in the range $0 < \gamma_o < 10^\circ$. In this range for  $\gamma_o$ three regions of distinct behavior occur (see Figs.~\ref{summary}a and d).  For small entry angles $0<\gamma_o<2.35^\circ$, duration of fillet contact  rises linearly (Fig.~\ref{summary}d) from 0 to $t_c = 0.075$ s but, because the normal curvature is  small even though it is negative, the gravity force is sufficient to maintain contact. The sled rejoins the floor when $v(t_c)=0$, simply riding over the low ``toe" of the fillet. For $2.35<\gamma_o<6.2^\circ$ the negative normal curvature and corresponding normal acceleration at the maximum value of fillet ``latitude" $v$ are too large to be supplied by the gravity force and contact is lost at increasing values of $v$ (Fig.~\ref{summary}a) as contact time gradually increases (Fig.~\ref{summary}d) from 0.030 to 0.063 s. For $\gamma_o>6.2^\circ$ contact persists all the way to the top of the fillet where $v_{max}=\pi/2$, but  contact time gradually decreases again to 0.031 s when $\gamma_o=10^\circ$.

The vertical component of velocity at loss of contact (Fig.~\ref{summary}b) rises more and more rapidly with increases in $\gamma_o$ and becomes a larger percentage of the lateral entry speed. When $\gamma_o=8^\circ$  the vertical velocity is  4.24 m/s leading to a zenith height of the resulting ballistic ejection trajectory of 1.04 m. Zenith height is extremely sensitive to entry angle. Increasing $\gamma_o$ from $8^\circ$ to $10^\circ$ doubles the zenith height of the flight path! 

The zenith occurs at a substantial distance $d_z$ from the point of loss of contact. Because the vertical component of velocity 4.24 m/s decreases the horizontal component only slightly more than 1\%, the horizontal distance $d_z$ to the zenith position is given approximately by

\begin{equation}
d_z=s_ot_f = s_o s_{lz} /g
\label{zenithdistance}
\end{equation}
where $t_f$ is the time of flight to the zenith. At a given entry speed $s_o$, both $t_f$ and $d_z$ vary roughly linearly with entry angle $\gamma_o$, but for fixed entry angle $\gamma_o$, $d_z$ varies approximately quadratically with initial speed $s_o$, as does $h_z$. Timing is important. For the example shown in Figs.~\ref{simresults} and~\ref{3Dcurve}, ($\gamma_o=8^\circ$), zenith height occurs at $t_f=0.43$ s and $d_z=17$ m. This is not far from the estimated diagonal cross track distance $d$ (see Table~\ref{Table1}) from the point of loss of inner wall contact to the exit wall. 

\section{Discussion} 

\subsection{Effect of Fillet on Track Ejection}

When the particle sled encounters the fillet at a non-zero entry angle $\gamma_o$, the lateral $y$ component of horizontal velocity  $s_{oy} = s_o\sin \gamma_o$ causes motion of the sled in the $v$ direction up the fillet. The fillet normal force resulting from its strong positive curvature acts perpendicular to this lateral velocity to turn it gradually upward while contact persists. If the inner wall were not curved but rather straight (as it is in straight sections), when the top of the fillet is reached the initially lateral component of velocity would be  transformed to vertical velocity of essentially the same magnitude.  The toroidal fillet acts less strongly (Fig.~\ref{summary}) because the surface is continuously turning away from its initial direction but it still is capable of generating vertical velocities and eventual free flight zenith heights larger than the outside exit wall heights intended and designed to contain the slider. 

It is a legitimate concern that the mass model used above is a particle while the actual slider and sled have  mass distributed over a lateral distance of the order of 0.5 m (the sled is restricted by rules~\cite{FIL2011} to a maximum width of 0.55 m). Another complication is that that the sled and slider composite body is not rigid but rather two bodies that can move relative to one another and can (and did) separate. Thus the results in Section III must be interpreted with care. Were the slider rigidly affixed to the sled then it is clear that an entry angle not much larger than $\gamma_o= 10^\circ $ would cause a vertical velocity of the right runner of substantially more than 6 m/s (Fig.~\ref{summary}b) and that roughly half of this would be contributed  to the center of mass (c.m.), sufficient for ejection. As the sled rides up the fillet, the horizontal component of the normal force increases and the sled lateral velocity decreases. Because the slider's body continues toward the wall over the right runner (Figs.~\ref{StillFrames}b,c), his c.m. will receive more than half the velocity of the runner, so that an estimate for the required entry angle for ejection of $\gamma_o> 11^\circ $ is probably not far off. It is also worth noting that the large horizontal component of the normal force acting on the front half of the sled late during contact accounts for the rotation, after loss of contact, of the slider and sled about the vertical axis in the direction observed in Fig.~\ref{StillFrames}c, d,e. 

It is essential to comment on the importance of the negative Gaussian curvature, i.e. the fillet at the base of an inside wall (see Appendix A), on the capacity for ejection. The positive lateral normal curvature of the fillet that was designed into it, $\kappa_2 =1/r$, gives it the capacity to turn the lateral component of incident velocity to a vertical one. The negative longitudinal normal curvature of the fillet and cylindrical wall above it, $\kappa_1=-\sin v /(R-r\sin v)$, means that contact must  eventually be lost at high speeds and large entry angles. Thereafter the subsequent center of mass flight path lies in a vertical plane essentially tangent to the inner wall with the near wall turning away from it and only the outside wall remaining to contain the slider in flight.  

Crucial structural components of luge sleds are two bridges~\cite{FIL2011}, essentially inverted-U-shaped symmetric steel beams (with approximate depth $\simeq 6$ mm and width $\simeq 50$ mm), that originate and terminate in the left and right runners, respectively, and pass through and support the sled body. The normal force as a function of time in the example calculations (Fig.~\ref{simresults}a) explains not only the vertical velocity required for ejection but also the deformation of the right sled runner and bridge observable in Figs.~\ref{StillFrames}d and e and claimed by the FIL Official Report~\cite{FIL2010} to be the cause of ejection. An analysis of the bridge (not included here) shows that bending deformations of this magnitude are reasonable given the magnitude ($\simeq14$ KN) of the average normal force later during contact. It also shows that early in contact the direction of the normal force was more nearly aligned with the lower part of the bridge near the runner contact point causing mostly axial compression along the bridge rather than bending. But the bridges are designed to bear loads in this direction.  Only later during the contact period is the normal force in a direction to cause bending. By this point most of the vertical velocity required for ejection has been attained (Fig.~\ref{simresults}b). 

Furthermore the catapulting (whatever this means precisely), if any occurred, happened late in the contact period when the result of catapulting would have been largely horizontal rather than vertical. By the time the normal force has the direction required for bending the bridge, the vertical velocity has largely been acquired. Catapulting did not produce the vertical velocity of ejection. Rather the vertical velocity of ejection is entirely accounted for by the very large normal forces early in the contact period due to the circular shape of the fillet.

The results above point incontrovertibly to the shape of the fillet as the proximate cause of ejection of Nodar Kumaritashvili from the Whistler Track. Contrary to the FIL report \cite{FIL2010} that the cause of the accident was a mysterious ``catapult effect", the sled was merely sliding on the ice surface in a way that was entirely predictable. The ice surface simply had the wrong shape and it was the interaction of the right runner with the fillet shape that caused the vertical velocity required for ejection - a velocity that was more than adequate to clear the exit wall. \textit{A fillet at the base of an inside wall can launch the slider into flight across the track.} 

If the fillet had not been present, and instead the wall and floor intersected at a right angle, the slider would have collided with the cylindrical wall with a horizontal lateral velocity of perhaps 4-7 m/s  (corresponding to an equivalent fall height of 0.8-2.5 m) onto the wall ice surface. The slider might have been severely bruised but would most likely still be alive. Flight sufficient to result in ejection would almost certainly not have occurred because the impulses from the cylindrical wall and the corresponding post-impact velocity would have been essentially horizontal.  

Yet track ejection consists of two essential components: 1) generation of velocity at loss of wall contact with magnitude and direction sufficient to result in flight, and 2) a failure to contain the resulting flight trajectory, if it occurs,  with sufficiently confining walls. Once flight takes place the surrounding safety surfaces must enclose the trajectory so that reentry from flight to sliding again occurs as a result of collision at low incidence angles with smooth, low-friction ice surfaces rather than at high angles with immovable fixed objects outside the enclosed tube. 

Safe design consists of first ensuring that no flight takes place, and secondly (even if the first goal is not achieved) that the path does not pass above and outside the enclosing walls. The Whistler track failed in both respects. In no case should a driving error by a slider have fatal consequences. As eloquently stated by Georgian President Mikahail Saakshvili, ``No sports mistake is supposed to lead to a human death."  Safe ice track design encompassing the above two aspects ensures this.

\subsection{Efficacy and Transparency of the Review and Investigation Process}

More than one year of effort by this author to contact and convince both the designer and Whistler Sport Legacies Society (WSLS) authorities to provide precise values for the  four geometric track parameters contained in Table~\ref{Table1} was completely unsuccessful. The designer's response to the initial request was that official approval was required from Whistler Sport Legacies Society, the present track owner. Numerous email and other messages over a period of six months to the WSLS authorities elicited similar replies: generally positive assurances that within a week or two the requested information could be made available. But nothing was ever provided. This is markedly different from the atmosphere at time of the last Canadian Olympics. The complete track design drawings (from which Figs.~\ref{crosssections}a and b are taken) were obtained in person from a government office in downtown Calgary in February 1988 in one hour.

New ice tracks are built every 4 years like clockwork. They are enormously expensive involving large sums of public money (the Whistler track cost C\$105M). Many of the parties involved potentially have financial interests that could conflict with their duty to put safety first. The FIL claims~\cite{FIL2010} it is committed ``to doing everything in its power to ensure ... that this tragic incident never happens again." Yet it is hard to believe that any independent, scientifically descriptive inquiry beyond the initial ``official" report~\cite{FIL2010} would have occurred, either by FIL or WSLS, without the recommendation of the Coroner~\cite{Coroner2010} that this happen. 
This entire process needs to be made more transparent and open, with the possibility of independent, financially disinterested, scientifically competent individuals involved at every stage and it shouldn't require a suggestion from a coroner to provide this. Details of the track design should be available to interested parties for the purpose of independent review and investigation rather than being guarded in a process that is as secret and closed as at present.

One is left to ponder why it has taken nearly two years to fully explain the cause of the accident, especially when the very day after it occurred the offending fillet was removed entirely from the track~\cite{FIL2010} as quoted in the second paragraph of this paper. The role of the FIL is particularly curious and even appalling. One wonders whom they thought would be convinced by the weak ``explanation" of the accident quoted in the introductory section above, suggesting that the circumstances of the accident were so complex and exceptional as to make it ``unknown and unpredictable"\cite{FIL2010}. The situation calls to mind the famous statement of Richard Feynman during the 1986 space shuttle Challenger investigation that, ``For a successful technology, reality must take precedence over public relations, for nature cannot be fooled"\cite{Gleick1992}.

It has been widely reported that the top speeds measured at the track are considerably (about 13\%) larger than the design values. Although the final design document "specified a maximum pre-calculated speed of 136.3 km/h [37.86 m/s]"~\cite{Coroner2010}, by 2009 "speeds of 154 km/h [42.78 m/s] had been reached," .... showing "that the designer's calculations of top speeds were incorrect."~\cite{Coroner2010}. The Coroner's report~\cite{Coroner2010} also describes fairly completely the history of the design process, the concerns of the FIBT and FIL, and the evolution of the realization of the excess speed problem.  All other things (track shape and entry angle in the collision) being equal, the lateral velocity is directly proportional to slider speed. A 13\% increase in total speed leads to a 13\% increase in lateral velocity, roughly a 13\% increase in vertical velocity at launch,  and a 27\% increase in peak height of the ballistic trajectory due to the vertical velocity. Thus, all other things being equal,  if the slider had been on a slower track traveling at a fractional speed 1/1.13 = 0.89  of the excessive speeds achieved on the Whistler track, his peak height would have been reduced by a factor of  1/1.27= 0.78 ((Fig.~\ref{summary}b) and he would have likely hit the inside top of the wall and been retained, rather than narrowly clearing it resulting in ejection. One also wonders whether the clearly perceived excess speed affected any design decisions, especially modification of exit wall heights as the design and speeds evolved before the accident.

\section{Summary and Conclusions} 

This rudimentary study has made clear how ice track ejection can be explained with a simple analytic model of the fillet surface shape and an equally simple model of dynamics on this surface based on Newton's laws. It nevertheless raises more questions than it answers. Some of these are:
1. Why is the fillet present in design specifications and drawings of curved inner wall shape designs when it is not required or even mentioned in the rules? What, if any, is its role?
2. Why, even if included at the base of the inner wall, was the specified value 0.125 m of fillet radius $r$ different from the only value of $r$, 0.1 m, mentioned in the rules? Are hand maintenance procedures able to adequately control satisfaction of this specification?
3. Were the exit wall heights adjusted throughout the evolution of the design after the actual speeds were realized to be larger than expected? If so, how? 
4. According to the FIL report, the ``existing safety wall, ... had already been lengthened and raised in the area of the accident,"\cite{FIL2010}.  On what basis was the height of wall calculated?
5. Why is the design process not more open and transparent, and more able to rely on the well-meaning scrutiny of interested but financially non-conflicted scientists and engineers around the world?

The main conclusions of this paper are:

1. The interaction of the right runner with the fillet at the bottom of the inner wall resulted in the vertical velocity necessary for, and thus was the cause of, the slider's ejection in the Vancouver Olympic luge accident.

2. The presence of fillets at the base of inner walls is an ice track design flaw that exposes sliders to ejection and should be prohibited in future designs. Furthermore these should be removed from all other turns at the Whistler track and other existing tracks. 

3. The bending of the bridge was certainly caused by the normal force from the fillet but this occurred during the last half of the contact period when the normal force was largely horizontal and thus its rebound was not able to provide vertical impulse.

4. A more open review and investigation process is highly desirable and could only increase the resulting safety of athletes using the tracks.

\section{Appendix A}  %

We begin with an analytic approximation for the toroidal fillet surface connecting the cylindrical inner wall and the planar floor and then present the equations of motion for a particle travelling on this specific surface, based on equations for motion of a particle on a general 2-D surface~\cite{Hubbard1989}. 

\subsubsection{Toroidal fillet geometry}

At the location of the Olympic accident the track slope $\beta=0$ and the fillet surface is well approximated by a torus (Theory similar to that presented in this paper can be used in the more general situation where $\beta\neq0$ and the fillet shape is a helical tube rather than a torus). When the floor slope angle $\beta = 0$, the implicit equation of the torus is  

\begin{equation}
(R-\sqrt{x^2+y^2})^2+z^2= r^2.
\label{torusimplicit}
\end{equation}

The $xyz$ coordinates of a point on the fillet are given by
\begin{align}
x(u,v) = (R-r\sin v)\sin u   \notag \\
y (u,v)= R(\cos u -1)-r\sin v\cos u   \\
z(u,v) = r(1-\cos v)   \notag
\label{fillet}
\end{align}

Equations 7 emphasize that the position vector to an arbitrary point on the torus is a function $\mathbf{r}=\mathbf{r}(u,v)$ of the two surface (angle) parameters $u$ and $v$~\cite{Faux1979}. The terminology of Faux and Pratt~\cite{Faux1979} is followed closely here.

The velocity vector lies along a tangent to the surface
\begin{equation}
{\mathbf{v}}=\dot{\mathbf{r}}= A\dot{\mathbf{u}}
\label{velocityvector}
\end{equation}
where
\begin{equation}
A=\left[
\begin{matrix}
\dfrac{\partial\mathbf{r}}{\partial u }& \dfrac{\partial\mathbf{r}}{\partial v} 
\end{matrix}
\right]
\end{equation}
and
\begin{equation}
\dot{\mathbf{u}}= \left[
\begin{matrix}
\dot{u} \\ \dot{v}
\end{matrix}
\right].
\end{equation}

The analytic expression for the torus surface makes it possible to compute analytically its partial derivatives with respect to $u$ and $v$. Specifically, 
\begin{equation}
\dfrac{\partial\mathbf{r}}{\partial u }=\left[
\begin{matrix}
(R-r\sin v)\cos u \\ -(R-r\sin v)\sin u  \\ 0
\end{matrix}
\right],
\end{equation}

\begin{equation}
\dfrac{\partial\mathbf{r}}{\partial v }=\left[
\begin{matrix}
-r\sin u\cos v \\ -r\cos u\cos v  \\ r\sin v
\end{matrix}
\right],
\end{equation}

\begin{equation}
\dfrac{\partial^2\mathbf{r}}{\partial u ^2}=\left[
\begin{matrix}
-(R-r\sin v)\sin u \\ -(R-r\sin v)\cos u  \\ 0
\end{matrix}
\right],
\end{equation}

\begin{equation}
\dfrac{\partial^2\mathbf{r}}{\partial v^2 }=\left[
\begin{matrix}
-r\sin u\sin v \\ -r\cos u\sin v  \\ r\cos v
\end{matrix}
\right],
\end{equation}

and
\begin{equation}
\dfrac{\partial^2\mathbf{r}}{\partial u \partial v}=\dfrac{\partial^2\mathbf{r}}{\partial v \partial u}=\left[
\begin{matrix}
-r\cos u\cos v \\ r\sin u\cos v  \\ 0
\end{matrix}
\right].
\end{equation}

The first and second fundamental matrices $\mathbf{G}$ and $\mathbf{D}$ of the toroidal surface are functions of the partial derivatives above:

\begin{equation}
\mathbf{G}=\mathbf{A^T A}=\left[
\begin{matrix}
\dfrac{\partial\mathbf{r}}{\partial u }\cdot
\dfrac{\partial\mathbf{r}}{\partial u } &\dfrac{\partial\mathbf{r}}{\partial u }\cdot
\dfrac{\partial\mathbf{r}}{\partial v }  \\   \\\dfrac{\partial\mathbf{r}}{\partial v }\cdot
\dfrac{\partial\mathbf{r}}{\partial u } &\dfrac{\partial\mathbf{r}}{\partial v }\cdot
\dfrac{\partial\mathbf{r}}{\partial v } 
\end{matrix}
\right]
=\left[
\begin{matrix}
(R-r\sin v)^2 & 0  \\ 0 &r^2
\end{matrix}
\right]
\end{equation}

and

\begin{equation}
\mathbf{D}=\left[
\begin{matrix}
\mathbf{n}\cdot\dfrac{\partial^2\mathbf{r}}{\partial u ^2} &\mathbf{n}\cdot\dfrac{\partial^2\mathbf{r}}{\partial u \partial v}
  \\ \mathbf{n}\cdot\dfrac{\partial^2\mathbf{r}}{\partial v \partial u} &\mathbf{n}\cdot
\dfrac{\partial^2\mathbf{r}}{\partial v^2 } 
\end{matrix}
\right]
=\left[
\begin{matrix}
-(R-r\sin v)\sin v &0
  \\ 0 &r
\end{matrix}
\right]
\end{equation}

and where the surface normal  $\mathbf{n}$ is given by 

\begin{equation}
\mathbf{n}=-(\dfrac{\partial\mathbf{r}}{\partial u }\times
\dfrac{\partial\mathbf{r}}{\partial v } )/|\dfrac{\partial\mathbf{r}}{\partial u }\times
\dfrac{\partial\mathbf{r}}{\partial v }|  
=\left[
\begin{matrix}
\sin v \sin u \\ \sin v \cos u  \\ \cos v
\end{matrix}
\right]
\label{normal}
\end{equation}
The negative sign in Eq.~\ref{normal} is chosen so that $\mathbf{n}$ points away from the ice surface and into the track (and the torus) interior. 

The normal curvature $\kappa_n$ of the torus in tangent  direction $\mathbf{v=A\dot u}$ is defined as the curvature of the curve of  intersection of the surface and the plane containing the tangent vector $\mathbf{v}$  and the surface normal $\mathbf{n}$~\cite{Faux1979}. It is determined by a ratio of quadratic forms in $\mathbf{\dot u}$ with the matrices $\mathbf{G}$ and $\mathbf{D}$ as coefficients~\cite{Faux1979}

\begin{equation}
\kappa_n=\dfrac{\mathbf{\dot u^T D \dot u}}{\mathbf{\dot u^T G \dot u}}.\label{normalcurvature}
\end{equation}

The directions in the surface at the point $(u,v)$ with respect to $\mathbf{\dot u}$ for which the normal curvature $\kappa_n$ becomes a minimum and maximum are called the principal directions of normal curvature, and the corresponding normal curvatures  $\kappa_1$ and $\kappa_2$ are called principal curvatures. The product of the principal curvatures, the Gaussian curvature  $K$, is also related to matrices $\mathbf{G}$ and $\mathbf{D}$ through
\begin{equation}
K=\kappa_1\kappa_2=\dfrac{\mathbf{|D|}}{\mathbf{|G|} }=
\dfrac{- \sin v}{r(R-r\sin v) }
\label{Gaussiancurvature}
\end{equation}

Due to symmetry, one of the principal curvatures of a torus is the inverse of the minor radius $\kappa_2=1/r$. Hence the other principal curvature is given by $\kappa_1=-\sin v /(R-r\sin v)$, the negative sign indicating that along the (longitudinal) first principal direction in question, the surface "falls away" from the inner pointing normal vector $\mathbf n$.

Normal curvature at a specific point $(u,v)$ varies continuously with direction $\phi$ according to Euler's formula~\cite{Gallier2002}
\begin{equation}
\kappa_n(\phi)=\kappa_1 \cos ^2 \phi + \kappa_2 \sin ^2 \phi
\label{normalcurvatureofphi}
\end{equation}
where $\phi$ is the angle measured in the tangent plane about $\mathbf n$ away from the first principal direction (of increasing "longitude") corresponding to curvature $\kappa_1$.

Because $R\gg r$, when the parameter $v>0$  the Gaussian curvature is negative and there is a direction in the surface tangent plane at critical angle $\phi_c$ for which  the normal curvature vanishes. This value of $\phi_c$ can then be calculated by substitution of $\kappa_n=0$ into Eq.~\ref{normalcurvatureofphi}.
\begin{equation}
\phi_c=\arctan{ \sqrt{ \dfrac{r\sin v}{R-r\sin v}}}.
\label{phizeroC}
\end{equation}
This direction separates those directions for which $\kappa_n>0$ and surface contact is maintained at all speeds from those in which $\kappa_n<0$ and loss of contact is possible at some speed.

\subsubsection{Equations of motion}

As the particle slides along the surface the parameters $u$ and $v$ and their derivatives change according to Newtons laws~\cite{Hubbard1989}. Here the generalized speeds~\cite{Kane1985}
are chosen to be the angular rates $\dot{u}$ and $\dot{v}$, the two components of $\mathbf{\dot u}$.

Although in general motion of luge, bobsled and skeleton on ice tracks all forces (including weight, normal force, aerodynamic lift and drag, ice friction and steering forces) are important for accurate prediction~\cite{Hubbard1989}, in the fillet contact problem considered herein the surface normal force resulting from normal acceleration $\kappa_n v^2$ due to fillet curvature is enormous. Further, the aerodynamic and friction forces are typically an order of magnitude smaller than the weight, the coefficient of  friction and drag area being of the order of 0.01 and 0.05 m$^2$, respectively~\cite{Mossner2011}~\cite{Balakin1988}. Thus only the first two of these forces, weight and normal force,  are included in the present analysis; the former only to ascertain its role in delaying loss of surface contact, and the latter because the vertical component of its impulse causes the vertical velocity required for ejection. 

Similar to the development in the Appendix of Ref.~\cite{Hubbard1989} the equations of motion for a particle sliding on the surface become (succinctly stated in terms of the matrices $\mathbf{G}$ and $\mathbf{D}$)
\begin{equation}
m\mathbf{G}\mathbf{\ddot u}=-mg
\left[
\begin{matrix}
\dfrac{\partial\mathbf{r}}{\partial u }\cdot \mathbf {k} \\ \dfrac{\partial\mathbf{r}}{\partial v} \cdot \mathbf {k} 
\end{matrix}
\right]-m\mathbf{\dot u^T D \dot u}
\label{EOM}
\end{equation}
where $\mathbf{k}$ denotes the vertical unit vector in the z direction. Note that Eq.~\ref{EOM} is independent of mass since each term contains the factor $m$.

The normal force $\mathbf{N}$ is a constraint force that insures that the luge remains in contact with the ice surface (when the normal curvature is positive). $N$ vanishes when the normal component of the weight $mg$ is insufficient to supply the negative normal acceleration (associated with negative normal curvature) to maintain surface contact. Loss of surface contact is synonymous with $\mathbf{N}$ vanishing. When $\mathbf{N}$ is nonzero its magnitude is given by 
\begin{equation}
N=m\kappa_n v^2 + mg\mathbf{n}\cdot\mathbf{k}.
\label{normalforce}
\end{equation}

\begin{acknowledgements}
The author acknowledges helpful discussions with Andy Ruina, Les Schaffer, Lyn Taylor, Frank Masley and Michael Drapack.
\end{acknowledgements}
\bibliographystyle{ieeetr}
\bibliography{aLUGE_References.bib}

\begin{thebibliography}{10}

\bibitem{FIL2010}
Anonymous, ``Official report to the {I}nternational {O}lympic {C}ommittee on
  the accident of {G}eorgian athlete, {N}odar {K}umaritashvili, at the
  {W}histler {S}liding {C}enter, {C}anada, on february 12, 2010, during
  official luge training for the {XXI} {O}lympic {W}inter {G}ames,'' tech.
  rep., International Luge Federation, 2010.
\newblock accessed December 2011.

\bibitem{Coroner2010}
T.~Pawlowski, ``Coroner's report into the death of {K}umaritashvili, {N}odar.''
  British Columbia Ministry of Public Safety and Solicitor General, 2010.
\newblock Case No: 2010-0269-0002.

\bibitem{FIL2011}
I.~L. Federation, {\em IRO International Luge Regulations - Artificial Track}.
\newblock International Luge Federation, 2010.

\bibitem{FIBT2011}
{International Bobsleigh and Skeleton Federation}, ``International {R}ules,
  {B}obsleigh.''
  \url{http://www.fibt.com/fileadmin/Rules/International_Rules_-_Bobsleigh_english_ver_1-1_16092007.pdf},
  2007.
\newblock accessed December 2011.

\bibitem{Hubbard1989}
M.~Hubbard, M.~Kallay, and P.~Rowhani, ``Three dimensional bobsled turning
  dynamics,'' {\em International Journal of Sports Biomechanics}, vol.~5,
  pp.~222--237, 1989.

\bibitem{Kelly2000}
A.~Kelly and M.~Hubbard, ``Design and construction of a bobsled driver training
  simulator,'' {\em Sports Engineering}, vol.~3, pp.~13--24, 2000.

\bibitem{Zhang1995}
Y.~L. Zhang, M.~Hubbard, and R.~K. Huffman, ``Optimum control of bobsled
  steering,'' {\em Journal of Optimization Theory and Applications}, vol.~85,
  pp.~1--19, 1995.

\bibitem{Faux1979}
I.~D. Faux and M.~J. Pratt, {\em Computational Geometry for Design and
  Manufacture}, pp.~105--113.
\newblock Chichester, UK: Ellis Horwood Ltd., 1979.

\bibitem{Gleick1992}
J.~Gleick, {\em Genius: The Life and Science of Richard Feynman}.
\newblock New York, NY: Vintage Books, 1st~ed., 1992.

\bibitem{Gallier2002}
J.~A. Gallier. \url{http://www.seas.upenn.edu/~cis70005}, 2002.
\newblock accessed December 2011.

\bibitem{Kane1985}
T.~R. Kane and D.~A. Levinson, {\em Dynamics: Theory and Applications}.
\newblock New York: McGraw-Hill, 1985.

\bibitem{Mossner2011}
M.~Mossner, M.~Hasler, K.~Schindelwig, P.~Kaps, and W.~Nachbauer, ``An
  approximate simulation model for initial luge track design,'' {\em J.
  Biomechanics}, vol.~44(5), 2011.

\bibitem{Balakin1988}
V.~A. Balakin, V.~N. Smirnov, and O.~V. Pereverzeva, ``Sliding of sled runner
  over ice,'' {\em Trenie i Iznos}, vol.~9(2), pp.~266--273, 1988.

\end{thebibliography}
\vfill\eject

\end{document}